\documentclass[aps,twocolumn,preprintnumbers,amsmath,amssymb,nofootinbib,superscriptaddress,notitlepage]{revtex4-1}

\usepackage{epsfig}
\usepackage[utf8]{inputenc}
\usepackage{braket}

\usepackage{color}
\usepackage{hyperref}
\usepackage{multirow}
\usepackage{tabularx}

\usepackage{slashed}
\usepackage{soul}
\usepackage{mathrsfs,bm}
\usepackage{txfonts}
\usepackage{indentfirst}
\usepackage{graphicx,booktabs}
\usepackage{color}
\usepackage{amssymb}
\usepackage{float}
\usepackage{subcaption}
\usepackage{ragged2e}

\usepackage{charter}

\begin{document}

\title{Constraining the hidden-charm pentaquark predictions and discriminating the $P_c(4440)$ and $P_c(4457)$ spins through the effective range expansion}

\author{Fang-Zheng Peng}

\affiliation{Southern Center for Nuclear-Science Theory (SCNT), Institute of Modern Physics, Chinese Academy of Sciences, Huizhou 516000, China}%

\author{Li-Sheng Geng}~\email{lisheng.geng@buaa.edu.cn}
\affiliation{School of Physics, Beihang University, Beijing 102206, China}%
\affiliation{Southern Center for Nuclear-Science Theory (SCNT), Institute of Modern Physics, Chinese Academy of Sciences, Huizhou 516000, China}%
\affiliation{Peng Huanwu Collaborative Center for Research and Education, Beihang University, Beijing 100191, China}
\affiliation{Beijing Key Laboratory of Advanced Nuclear Materials and Physics, Beihang University, Beijing 102206, China}

\author{Ju-Jun Xie}~\email{xiejujun@impcas.ac.cn}

\affiliation{Southern Center for Nuclear-Science Theory (SCNT), Institute of Modern Physics, Chinese Academy of Sciences, Huizhou 516000, China}%
\affiliation{Heavy Ion Science and Technology Key Laboratory, Institute of Modern Physics, Chinese Academy of Sciences, Lanzhou 730000, China}%
\affiliation{School of Nuclear Sciences and Technology, University of Chinese Academy of Sciences, Beijing 101408, China}

\date{\today}

\begin{abstract}

The Weinberg compositeness criterion dictates that a pure shallow bound state is characterized by a large scattering length $a_0\gg\mathcal{O}(1/\beta)$ and a positive effective range $r_0$ that naturally scales to the size of $\mathcal{O}(1/\beta)$, where $1/\beta$ signifies the interaction range. In constructing the contact-range effective field theory (EFT) up to the next-to-leading order to describe the pentaquarks  $P_c(4312)$, $P_c(4440)$, and $P_c(4457)$ observed by the LHCb collaboration in 2019, we match the effective range $r_0$ at single-channel situation for these pentaquarks with the low-energy couplings within the EFT framework. Three different schemes are used to connect the couplings with the effective range. We find positive effective ranges $r_0$ of the natural size of $\mathcal{O}(1/\beta)$  for the spin configurations $J^P=\frac{3}{2}^-$ for  $P_c(4440)$ and $J^P=\frac{1}{2}^-$ for $P_c(4457)$ within the molecular $\bar{D}^* \Sigma_c$ description. Additionally, predictions from the power counting for low-energy couplings or Wilsonian coefficients suggest that, under heavy quark spin symmetry, the broad $P_c(4380)$ resonance, discovered by the LHCb collaboration in 2015, when considered as part of the single-channel $\bar{D}^{(*)} \Sigma_c^{(*)}$ molecular system alongside $P_c(4312)$, $P_c(4440)$, and $P_c(4457)$, has a mass of approximately $4376$ $\rm{MeV}$. 

\end{abstract}

\maketitle

\section{Introduction}

The theoretical and experimental investigation of multiquark exotic hadrons, which transcend the conventional quark model description, has consistently triggered substantial interest due to their potential implications for our understanding of nonperturbative quantum chromodynamics~\cite{Gell-Mann:1964ewy,Zweig:1964ruk,Brambilla:2010cs,Esposito:2016noz,Ali:2017jda}. In 2015, two hidden-charm pentaquark states, $P_{c}(4380)$ and $P_{c}(4450)$, were firstly discovered by the LHCb experiment~\cite{LHCb:2015yax}. Subsequent observations revealed that the $P_{c}(4450)$ splits into two narrow peaks, namely, $P_c(4440)$ and $P_c(4457)$, and a new $P_c(4312)$ state was also identified. However, the broad $P_{c}(4380)$ does not appear as a clear signal in the $J/\psi p$ mass spectrum~\cite{LHCb:2019-pentaquarks}. These three pentaquarks, $P_c(4312)$ ($P_{c1}$), $P_c(4440)$ ($P_{c2}$), and $P_c(4457)$ ($P_{c3}$), have sparked lively discussions regarding their nature~\cite{Chen:2016qju,Guo:2017jvc,Fernandez-Ramirez:2019koa,Liu:2019zoy,Guo:2019fdo,Xiao:2019aya,Wang:2019nwt,Meng:2019ilv,Wu:2019adv,Voloshin:2019aut,Burns:2019iih}. Their masses and widths are
\begin{footnotesize}
\begin{eqnarray}
    m(P_{c1}) &=& 4311.9\pm0.7^{+6.8}_{-0.6} \, \mathrm{MeV} \, , \ \ \Gamma(P_{c1}) = 9.8\pm2.7^{+3.7}_{-4.5} \, \mathrm{MeV} , \\
    m(P_{c2}) &=& 4440.3\pm1.3^{+4.1}_{-4.7} \, \mathrm{MeV} \, , \ \ \Gamma(P_{c2}) = 20.6\pm4.9^{+8.7}_{-10.1} \, \mathrm{MeV} ,\\
    m(P_{c3}) &=& 4457.3\pm0.6^{+4.1}_{-1.7} \, \mathrm{MeV} \, , \ \ \Gamma(P_{c3}) = 6.4\pm2.0^{+5.7}_{-1.9} \, \mathrm{MeV} .
\end{eqnarray}
\end{footnotesize}
Since their masses are close to the $\bar{D} \Sigma_c$ and $\bar{D}^* \Sigma_c$ thresholds, it is natural to consider them as molecules of charmed mesons and  baryons~\cite{Wu:2010jy,Wang:2011rga,Yang:2011wz,Yuan:2012wz,Wu:2012md,Xiao:2013yca,Uchino:2015uha,Karliner:2015ina,Liu:2018zzu,Guo:2019kdc,Sakai:2019qph,Peng:2020gwk,Du:2021fmf,Peng:2022iez,Liu:2024uxn}. Besides, there exist other explanations for these pentaquarks, e.g., compact five quark states and hadrocharmonia~\cite{Zhu:2019iwm,Ali:2019npk,Eides:2019tgv,Giron:2019bcs,Kuang:2020bnk,Garcilazo:2022kra}.

Within the molecular picture, heavy quark spin symmetry (HQSS) implies other four pentaquarks formed by $\bar{D} \Sigma_c^*$ and $\bar{D}^* \Sigma_c^*$ ~\cite{Liu:2019-prl,Du:2019pij}, which together with the above three pentaquarks already identified by the LHCb collaboration, constitute a multiplet of seven pentaquark states. Furthermore, within this hadronic molecular framework, the mass spectrum alone derived from the effective field theory or phenomenology does not allow for a definitive determination of the $P_c(4440)$ and $P_c(4457)$ spins, which can be either $J^P=\frac{1}{2}^-$ or $\frac{3}{2}^-$. Many studies have addressed this spin issue~\cite{PavonValderrama:2019nbk,Chen:2019asm,Liu:2023wfo,Liu:2019zvb,Zhang:2023czx}. To determine the $P_{c2}$ and $P_{c3}$ spins under the two-body hadronic molecular $\bar{D}^* \Sigma_c$ description, it is helpful to reconsider the nature of being ``molecule" or ``composite". For a given few-body system on a low energy scale, the Weinberg compositeness criterion could judge whether this system is ``composite" or more like ``elementary". The Weinberg compositeness criterion was originally proposed for studying the deuteron to discriminate between the elementary particle $\ket{d}$ and the composite molecular state $\ket{np}$, and this criterion has been extended to exotic hadronic states. A multiquark state $\ket{\Phi}$ can contain two components, a compact one $\ket{\phi}$ and a two-body molecular part $\ket{h_1h_2(\bold{k})}$
\begin{eqnarray}
\ket{\Phi}=\sqrt{Z}\ket{\phi}+\int\frac{d^3\bold{k}}{(2\pi)^3}\lambda(\bold{k})\ket{h_1h_2(\bold{k})} ,
\end{eqnarray} 
with $h_i$ (i=1,2) representing the hadrons in the molecule picture.  $Z=|\langle\Phi|\phi\rangle|^2$ and $1-Z=\int\frac{d^3\bold{k}}{(2\pi)^3}|\lambda(\bold{k})|^2$ indicate the probability of the compact and molecular parts within the given system, respectively.

For an $s$-wave loosely bound state, the Weinberg compositeness criterion connects the renormalization factor $Z$ to the scattering length $a$ and effective range $r$. For a pure molecular state, the derivation above yields the condition $Z=0$, which implies that the scattering length $a$ should be \textit{unnaturally} large, namely, $a\gg 1/\Lambda$ with $\Lambda$ the hard energy scale of the molecular system, while the effective range $r$ is positive and of the order of the range $\mathcal{R}$ of the potential responsible for binding the molecular state.

This study aims to investigate the effective ranges of the $P_{c}(4312)$, $P_{c}(4440)$, and $P_{c}(4457)$ states within the molecular picture and the contact effective field theory framework. We will consider two alternative scenarios with distinct $\bar{D}^* \Sigma_c$ spin configurations for $P_{c}(4440)$ and $P_{c}(4457)$ to calculate the effective range. By evaluating which scenario yields a \textit{unnaturally} large scattering length $a$ and a positive effective range $r$ comparable to the interaction range $\mathcal{R}$ --- consistent with the molecular hypothesis with $Z=0$ --- we aim to determine the spins of $P_{c}(4440)$ and $P_{c}(4457)$. Besides, the existence of other predicted $P_c$ molecular states under HQSS will be examined, which is important for future experiments and understanding the pentaquarks' nature.

This article is structured as follows. In Sect.~\ref{sec: 2-ERE and compositeness}, we derive the effective range and scattering length for pure molecules by revisiting the Weinberg compositeness. In Sect.~\ref{sec: 3-NLO Contact field theory}, we construct the next-to-leading order contact field theory to describe the $P_c$ states within the molecular picture. The corresponding power counting is introduced in Sect.~\ref{sec: 4-power counting}. Three matching schemes to relate the effective range to low-energy couplings are proposed in Sect.~\ref{sec: 5-match r with V}. The numerical results are presented in Sect.~\ref{sec: 6-results}. Finally, Sect.~\ref{sec: 7-conclusions} provides our conclusions and discussions.
 
\section{Effective range expansion and the Weinberg compositeness of bound states} \label{sec: 2-ERE and compositeness}

Considering the $P_c$ states as two-body bound system of $\bar{D}^{(*)}\Sigma_c^{(*)}$, the effective range expansion at low momenta $p$ yields \cite{Hammer:2019-Nuclear-EFT, Contessi:2024vae} 
\begin{eqnarray}
   p^{2l+1} \text{cot}\delta_l=-\frac{1}{a_l}+\frac{1}{2}r_l p^2+\cdots, 
\end{eqnarray}
where $l=1,2,3, \cdots$ represent the different partial waves, $\delta_l$ is the phase shift, $a_l$ and $r_l$ are respectively the scattering length and effective range. 
The $s$-wave amplitude in a low-energy scattering process can then be written 
\begin{eqnarray}
    \mathcal{T}=\frac{2\pi}{\mu}\frac{1}{p \; \text{cot}\delta_0-ip}=\frac{2\pi}{\mu}\frac{1}{-\frac{1}{a_0}+\frac{1}{2}r_0 p^2+\cdots} \, ,
    \label{eq:amplitude with ERE}
\end{eqnarray}
with $\mu$ denoting the reduced mass of the two-body bound system.

The binding momentum for the $\bar{D}^{(*)}\Sigma_c^{(*)}$ molecules is $\gamma=\sqrt{2\mu E_B}$, where $E_B>0$ is the binding energy. For loosely bound states, $\gamma \ll \beta$, where $\beta$ is the hard energy scale of the two-body system, usually represented by the mass of the exchanged meson $m_{ex}$. The compositeness $1-Z$ under the non-relativistic situation could be written as the following \cite{Weinberg:1962hj,Weinberg:1965zz}
\begin{eqnarray}
    1-Z &=& \int\frac{d^3\bold{k}}{(2\pi)^3}\frac{\bra{h_1h_2(\bold{k})}V\ket{\Phi}}{(E_B+\frac{\bold{k}^2}{2\mu})^2}(1+\mathcal{O}(\frac{\bold{k}^2}{{\beta}^2}))  \nonumber \\
    &=& \frac{\mu^2 g^2}{2\pi\gamma}(1+\mathcal{O}(\frac{\gamma}{{\beta}})) \, ,
\end{eqnarray} 
where $V$ is the interaction between the state $\ket{\Phi}$ and the hadron pair, and the constant $g\simeq\bra{h_1h_2(\bold{k})}V\ket{\Phi}$ introduced here is the coupling between the molecular system and the given state. Through the dispersion relation of $\mathcal{T}$, this coupling $g$ can be connected to the scattering amplitude. Comparing with $\mathcal{T}$ in Eq.~(\ref{eq:amplitude with ERE}), the Weinberg compositeness could be obtained as \cite{vanKolck:2022lqz,Baru:2021ldu}
\begin{eqnarray}
    a_0 &=& 2\frac{(1-Z)}{\gamma (2-Z)} + \mathcal{O}(\frac{1}{m_{ex}}) 
    \label{eq:Weinberg compositeness for a_0} \, ,\\
    r_0 &=& -\frac{Z}{\gamma (1-Z)}+ \mathcal{O}(\frac{1}{m_{ex}}) 
    \label{eq:Weinberg compositeness} \, .
\end{eqnarray}

We note that the compositeness $1-Z$  is very sensitive to $a_0$ and $r_0$, because the compositeness $1-Z=\sqrt{a_0/(a_0-2r_0)}$ becomes singular at $r_0=a_0/2$. On the other hand, the effective range $r_0$ and scattering length $a_0$ of the deuteron can be well described by assuming $Z=0$~\cite{Weinberg:1965zz}. In this work, we assume that the above relations hold. More extensive studies on the limitations and extensions of the Weinberg compositeness criterion, including scenarios involving virtual states, resonant states and coupled-channel scenario, can be found in Refs.~\cite{Hyodo:2013nka,Hyodo:2013iga,Oller:2017alp,Matuschek:2020gqe,Baru:2021ldu,Li:2021cue,Song:2022yvz,Kinugawa:2024crb}.

According to the above relations among $a_0$, $r_0$, and the compositeness $1-Z$, for a pure molecular system with $Z=0$, the effective range $r_0$ is positive~\cite{Landau:1991wop,Esposito:2021vhu, Bethe:1949yr}, with a value of the order of the interaction range of $\mathcal{O}(1/m_{ex}) \sim \mathcal{R}$. Conversely, for a compact state with $Z=1$, $r_0$ is negative. Additionally, under the low-energy situation of $\gamma \ll m_{ex}$, assuming $Z=0$ for a molecule leads to an \textit{unnaturally} large scattering length $a_0 \sim 1/\gamma \gg \mathcal{R}$. This is consistent with the characteristics of shallow bound states, as the scattering length $a_0$ could be interpreted as the node of the wave function $\phi(x)$ at low energy $k$ \cite{Sakurai:2011zz}
 \begin{eqnarray}
     \phi(x) \sim \text{sin} \ (k x +\delta_0) \sim 1-\frac{x}{a_0} \, .
 \end{eqnarray}
The emergence of a shallow bound state will alter $a_0$ from infinity to an $\textit{unnaturally}$ large magnitude. 

Therefore, the molecular hypothesis for the pentaquark states as $\bar{D}^{(*)}\Sigma_c^{(*)}$ bound states imply positive natural effective ranges $r_0 \sim \mathcal{R}$ and unnaturally large scattering lengths of $a_0 \gg \mathcal{R}$ for
the $\bar{D}^{(*)}\Sigma_c^{(*)}$ interactions. However, it should be noted here that the above conclusion is for \textit{bound states} since the application of the Weinberg compositeness criterion to virtual or resonance states is different~\cite{Hyodo:2013iga,Matuschek:2020gqe,Li:2021cue,Meng:2022ozq,Shen:2024npc}. In addition, the assumption of \textit{pure molecule states} is necessary for obtaining a positive effective range $r_0$. This is related to the uncertainties from $\mathcal{O}(1/m_{ex})$ and can be understood as follows. Taking, as an example, the requirement of $r_0>0$ in Eq.~\eqref{eq:Weinberg compositeness}, we have 
\begin{eqnarray}
    \frac{1}{m_{ex}} < \frac{1-Z}{Z}\gamma \, .
    \label{eq. Z value limitted}
\end{eqnarray}
For the $P_c$ states studied here, one can assign $m_{ex}$ a value of $\sim 500 \, \mathrm{MeV}$, assuming that the short-range interactions are mediated by the scalar meson of $\sigma$ and vector mesons of $\rho$ and $\omega$. On the other hand, $\gamma$ is taken as about $100 \, \mathrm{MeV}$~\cite{Yan:2023ttx, Peng:2021hkr, Peng:2019wys}.
As a result, a compact component of $Z\geq \, 0.2$ will invalidate the relation of Eq.~\eqref{eq. Z value limitted} and thus the conclusion of $r_0>0$ for bound molecular states. Consequently, the premise of pure molecular states or a negligible fraction of compact components for the $P_c$ states is essential for the above conclusions.

\section{Molecular descriptions of $P_c$ states with contact effective field theory up to Next-to Leading Order} \label{sec: 3-NLO Contact field theory}

The $P_c(4312)$, $P_c(4440)$, and $P_c(4457)$ states can be described as non-relativistic $\bar{D}^{(*)}\Sigma_c$  bound states in the contact effective field theory \cite{Liu:2018zzu,Liu:2019-prl,Peng:2020-Pcs-4459,Yan:2022wuz-Pcs-4338}. Under HQSS, the LO meson-baryon potential can be expressed with two low-energy couplings 
\begin{eqnarray}
    V=C_a+ C_b (\Vec{\sigma}_{1L} \cdot \Vec{S}_{2L}),
\end{eqnarray}
where $\Vec{\sigma}_{1L}$ and $\Vec{S}_{2L}$ represent the spins of the light degrees of freedom within the charmed mesons and baryons, respectively. According to HQSS, the spins of heavy $c$ quarks do not influence the interactions between the $\bar{D}^{(*)}$ mesons and $\Sigma_c^{(*)}$ baryons. Here, it should be noted that our analysis assumes a single-channel of $\bar{D}^{(*)}\Sigma_c$ configuration, where we study with the isospin-symmetric limit and use the isospin-averaged masses of $\bar{D}^{(*)}$ and $\Sigma_c$ as listed in the Particle Data Group (PDG)~\cite{ParticleDataGroup:2024cfk}. The above contact potential yields the following
potentials for the $\bar{D}\Sigma_c$ and $\bar{D}^{*}\Sigma_c$ systems
\begin{eqnarray}
    V(\bar{D}\Sigma_c \, ; J^P={\frac{1}{2}}^- ) &=& C_a \, , \\
    V(\bar{D}^{*}\Sigma_c \, ; J^P={\frac{1}{2}}^-) &=& C_a-\frac{4}{3}C_b \, ,\\
    V(\bar{D}^{*}\Sigma_c \, ; J^P={\frac{3}{2}}^-) &=& C_a+\frac{2}{3}C_b \, .
\end{eqnarray}
However, the spin classification of $\bar{D}^{*}\Sigma_c$ for $P_c(4440)$ and $P_c(4457)$ remains undetermined from the mass spectra studies with the above LO potential. $P_c(4440)$ and $P_c(4457)$ can either have $J^P=\frac{1}{2}^-$ or $\frac{3}{2}^-$. Both scenarios predict a $\bar{D}\Sigma_c$ bound state, corresponding to $P_c(4312)$, of a similar mass within the experimental uncertainty.

Since the effective range $r_0$ should be positive and of natural size for pure $s$-wave bound molecular states, it is useful to calculate $r_0$ from the contact range EFT to help determine which spin scenario for $P_c(4440)$ and $P_c(4457)$ is preferred. In the scenario where the scattering length $a_0 \sim 1/p \gg \mathcal{R}$ is unnaturally large, to generate the effective range $r_0$, one needs to construct the contact potential at least up to the next-to-leading order (NLO) of $\mathcal{O}(p^2)$. This can be easily verified by expanding the amplitude of Eq.(6)~\cite{Kaplan:1998tg,Kaplan:1998we}. 

In the non-relativistic effective field theory, the contact interactions between two hadrons are described by the following effective Lagrangian~\cite{Hammer:2000xg, Hammer:2019-Nuclear-EFT}
\begin{eqnarray}
\mathcal{L} &=& \psi^\dagger(i\partial_t+\frac{\nabla^2}{2m})\psi+C_0\psi^{\dagger}\psi\psi^{\dagger}\psi \nonumber \\
&+& C_2[(\psi^{\dagger}\psi^{\dagger})(\psi\overset{\leftrightarrow}{\nabla}^{2}\psi)+H.c.]+... \, .
\label{eq:non-relativistic contact EFT}
\end{eqnarray}
From the above Lagrangian, the contact potential $V$ in the center-of-mass system (CMS) can be derived as \cite{Ordonez:1995rz,Epelbaum:1998ka,Epelbaum:2004fk,Haidenbauer:2013oca}
\begin{eqnarray}
  V &=& C_0 + C_2^1q^2 + C_2^2k^2 + (C_2^3q^2+ C_2^4k^2)\Vec{\sigma}_{1L}\cdot\Vec{S}_{2L} \nonumber \\ && + \frac{i}{2}C_2^5(\Vec{\sigma}_{1L}+\Vec{S}_{2L})\cdot(q\times k) + C_2^6(q\cdot\Vec{\sigma}_{1L})(q\cdot\Vec{S}_{2L}) \nonumber \\ 
   && + C_2^7(k\cdot\Vec{\sigma}_{1L})(k\cdot\Vec{S}_{2L}) + \frac{i}{2}C_2^8(\Vec{\sigma}_{1L}-\Vec{S}_{2L})\cdot(q\times k) \nonumber \\ 
   && + \cdots \, ,
    \label{eq:contact potential}
\end{eqnarray}
where $\Vec{p}$ and $\Vec{p'}$ are the initial and final center-of-mass momenta,  $C_0$, $C_2^{i}$ ($i=1,\cdots, 8$) are the low-energy couplings, also known as Wilson coefficients in effective field theory, and $q=\Vec{p'}-\Vec{p}$, $k=(\Vec{p'}+\Vec{p})/2$ are defined as the transferred and average momenta.

Performing the standard partial-wave projection, the above NLO contact potential can be respectively reduced to the following forms with four low-energy couplings for the $s$-wave states of $^{2}S_{\frac{1}{2}}$ $\bar{D}\Sigma_c$, $^{2}S_{\frac{1}{2}}$ $\bar{D}^{*}\Sigma_c$, and $^{4}S_{\frac{3}{2}}$ $\bar{D}^{*}\Sigma_c$
\begin{eqnarray}
    V_1 &=& C_a + \frac{1}{2}D_a(\Vec{p}^2 + \Vec{p'}^{2}) 
    \label{eq: contact potential for Pcs 1} \\
    V_2 &=& C_a-\frac{4}{3}C_b+\frac{1}{2}(D_a-D_b)(\Vec{p}^2 + \Vec{p'}^{2}) 
    \label{eq: contact potential for Pcs 2} \\
    V_3 &=& C_a+\frac{2}{3}C_b+\frac{1}{2}(D_a+\frac{D_b}{2})(\Vec{p}^2 + \Vec{p'}^{2}) \, .
    \label{eq: contact potential for Pcs 3}
\end{eqnarray}
 The Lippmann-Schwinger equation of $\mathcal{T}=V+VG\mathcal{T}$ can be greatly simplified in the contact effective field theory.
 Introducing a Gaussian regulator of the form $e^{-\frac{2 \, q^2}{\Lambda^2}}$ with $\Lambda$ the momentum cutoff and adopting the
 on-shell approximation, the scattering amplitude $\mathcal{T}$  becomes
\begin{eqnarray}
   \mathcal{T}&=&V[1+V \, G(q,\Lambda)+(V \, G(q,\Lambda))^2+\cdots] \nonumber \\
   &=&\frac{V}{1-VG(q,\Lambda)}
   \label{eq. T=V+VGT},
\end{eqnarray}
where the propagator $G$ is
\begin{eqnarray}
    G(E_B,\Lambda)=\frac{\mu}{\pi^2}\int_0^{+\infty}dq\frac{q^2}{2\mu \, E_B+q^2}\, e^{-\frac{2 \, q^2}{\Lambda^2}} \, .
\end{eqnarray}
Here, $\mu$ is the reduced mass of the $\bar{D}^{(*)}\Sigma_c$ system and $E_B>0$ is the binding energy.

\begin{figure}[htbp]
  \centering
   \includegraphics[scale=0.53]{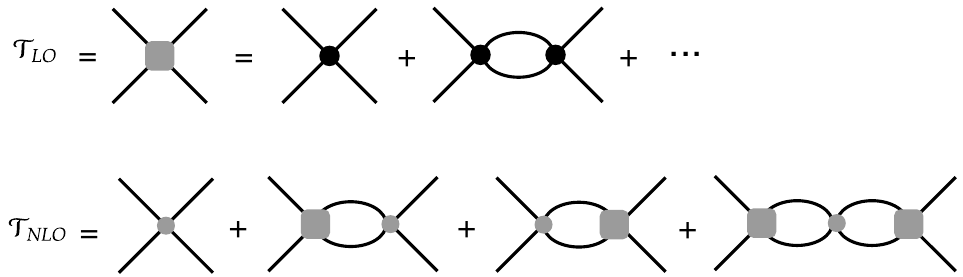}
  \caption{\justifying{The $s$-wave amplitudes of LO and NLO terms from the contact interactions, where black and gray solid circles denote the $C_0$ and $C_2$  terms.}}    \label{fig:T-amplititudes}
\end{figure}

In the unnatural scenario where the scattering length $a_0$ is large, satisfying $a_0 \sim \frac{1}{p} \gg \frac{1}{\Lambda}$ [corresponding precisely to our case with $Z=0$ in Eq.~\eqref{eq:Weinberg compositeness for a_0}], the amplitude $\mathcal{T}$ can be expanded in powers of $\frac{p}{\Lambda}$ rather than $p$. This expansion is permissible for large $a_0$ situation, because $p$ is not small compared to $\frac{1}{a_0}$, thereby leading to~\cite{Kaplan:1998tg,Kaplan:1998we,Hammer:2019-Nuclear-EFT}
\begin{eqnarray}
     \mathcal{T}=-\frac{2\pi}{\mu}\frac{1}{\frac{1}{a_0}+ip}[1+\frac{r_0}{2(\frac{1}{a_0}+ip)}p^2 +\cdots] \, .
\end{eqnarray} 
Matching the above equation to the LO and NLO terms, $\mathcal{T}_{LO}$ and $\mathcal{T}_{NLO}$, from Eq.~\eqref{eq. T=V+VGT}, as illustrated in Fig.~\ref{fig:T-amplititudes}, one can derive the relations among the scattering length $a_0$, effective range $r_0$, and low-energy couplings $C_0$ and $C_2$  as \cite{Cohen:1996my,Epelbaum:2010nr,Kaplan:1998tg,Kaplan:1998we}
\begin{eqnarray}
    C_0 &=& \frac{2\pi}{\mu}\frac{1}{\frac{1}{a_0}-\alpha\Lambda} \, , \label{eq:a0 and r0 as couplings-1} \\ 
    C_2&=& \frac{2\pi}{\mu}\frac{1}{(\frac{1}{a_0}-\alpha\Lambda)^2}\frac{r_0}{2} \, ,
    \label{eq:a0 and r0 as couplings}
\end{eqnarray}
where $\alpha$ can be determined with different regulators and renormalization schemes~\cite{Hammer:2019-Nuclear-EFT}. For the Gaussian form factor we used here, $\alpha\Lambda = \frac{2\pi}{\mu}G(0,\Lambda)\sim 0.4 \ \Lambda$. 

\section{Power counting analysis} \label{sec: 4-power counting}

Two power counting schemes --- the Weinberg power counting and the KSW (Kaplan, Savage and Wise) power counting --- can both be applied to the above low-energy couplings $C_{0}$ and $C_{2}$ to ensure that the LO and NLO terms of the amplitude $\mathcal{T}$ are matched to the resummation diagrams depicted in Fig.~\ref{fig:T-amplititudes} \cite{Epelbaum:2017byx}. 

Considering a given Effective Field Theory for a particular system, the effective Lagrangian density can be expressed in the following form, based on the Operator Product Expansion (OPE)
  \begin{eqnarray}
      \mathcal{L}_{eff}=\sum_n \, C_n \, \mathcal{O}_n \, ,
  \end{eqnarray}
where $\mathcal{O}_n$ are the operators that satisfy the symmetry requirements for constructing the EFT, and $C_n$ stands for the corresponding Wilsonian coefficient. 
In this EFT, employing the Wilsonian cutoff $\Lambda$, the naive dimensional analysis (NDA) dictates a condition for naturalness, suggesting that the dimensionless parameters in OPE should be of $\mathcal{O}(1)$. Consequently, the different Wilsonian coefficients $C_i$ and $C_j$ associated with the operators $\mathcal{O}_i$ and $\mathcal{O}_j$ are related
 \begin{eqnarray}
     \frac{C_i}{C_j}\sim\frac{\Lambda^{d_j}}{\Lambda^{d_i}} \, ,
 \end{eqnarray}
where $d_i$ and $d_j$ are the mass dimensions of the operators $\mathcal{O}_i$ and $\mathcal{O}_j$.   
This relation indicates that for the operators $\mathcal{O}_n$ with higher mass dimensions, the corresponding Wilsonian coefficients $C_n$ will be suppressed compared to those with lower mass dimensions.
The above conclusion regarding Wilsonian coefficients from NDA is consistent with Weinberg's power counting used for the contact EFT in Eq.~(\ref{eq:non-relativistic contact EFT}), describing the $\bar{D}^{(*)}\Sigma_c^{(*)}$ interactions, where the ratio between $C_2$ and $C_0$ is derived as 
\begin{eqnarray}
    \frac{C_2}{C_0}\sim\Lambda^{-2} \, ,
    \label{eq: R-value}
\end{eqnarray}
since $C_2$ is of $\mathcal{O}(p^2)$ while $C_0$ is of $\mathcal{O}(p^0)$.

Introducing the scheme of \textit{power divergence subtraction} (PDS) in dimensional regularization, Kaplan, Savage, and Wise \cite{Kaplan:1998tg,Kaplan:1998we} suggest that for the large scattering length situation, the couplings $C_0$ and $C_2$ in Eq.~\eqref{eq:a0 and r0 as couplings-1} and \eqref{eq:a0 and r0 as couplings} will scale as 
 \begin{eqnarray}
     \frac{C_2}{C_0}\sim\frac{1}{\Lambda(\alpha\Lambda)} \, .
     \label{eq:KSW power counting}
 \end{eqnarray}
It is clear that with $\alpha\Lambda\rightarrow\Lambda$, Weinberg's power counting in Eq.~(\ref{eq: R-value}) will be recovered, while the choice of $\alpha\Lambda\rightarrow p$ leads to the KSW counting \cite{Epelbaum:2017byx}.

\section{Matching the effective range with the NLO contact potential} \label{sec: 5-match r with V}
To determine the effective range $r_0$, it is necessary to include the NLO momentum-related term of the contact potential, as given in Eqs.~\eqref{eq: contact potential for Pcs 1}-\eqref{eq: contact potential for Pcs 3}, to describe the  $P_c(4312)$, $P_c(4440)$, and $P_c(4457)$ pentaquarks. With the on-shell approximation, the $\bar{D}^{(*)}\Sigma_c$ potentials can be expressed as 
\begin{eqnarray}
    V(\bar{D}\Sigma_c) &=& C_a+D_a p_{cm1}^2 \, , \label{eq: NLO-1}
    \\
    V(\bar{D}^{*}\Sigma_c\, , {\frac{1}{2}}^-) &=& C_a-\frac{4}{3}C_b+(D_a-D_b) p_{cm2}^2 \, , \label{eq: NLO-2} 
    \\
    V(\bar{D}^{*}\Sigma_c \, , {\frac{3}{2}}^-) &=& C_a+\frac{2}{3}C_b+(D_a+\frac{D_b}{2}) p_{cm2}^2 
    \label{eq: NLO-3}  
\end{eqnarray}
where $p_{cm1}$ and $p_{cm2}$ are the center-of-mass momenta for $\bar{D}\Sigma_c$ and $\bar{D}^{*}\Sigma_c$, respectively, given by
\begin{eqnarray}
    p_{cm}=\frac{\sqrt{[s-{(m_1 + m_2)}^2][s-{(m_1 - m_2)}^2]}}{2\sqrt{s}} \, .
\end{eqnarray}
Here, $\sqrt{s}$ represents the center-of-mass energy, $m_1$ and $m_2$ denote the masses of $\bar{D}^{(*)}$ and $\Sigma_c$. However, because there are only three experimental masses, it is impossible to predict the effective ranges from the potential given in Eq.~(\ref{eq: NLO-1})-\eqref{eq: NLO-3} with four low-energy couplings, i.e., $C_a$, $C_b$, $D_a$ and $D_b$. Therefore, we propose three schemes to address this issue:
\begin{itemize}
\item[(\uppercase\expandafter{\romannumeral1})] In Scheme I, we neglect the spin-spin interaction term at $\mathcal{O}(p_{cm}^2)$ of the potential in Eqs.~(\ref{eq: NLO-1})-\eqref{eq: NLO-3} , namely, setting $D_b=0$. This approach is reasonable because the difference in the $C_2$ term is of the next-to-leading order (NLO) and should contribute less than other terms \cite{Wu:2024bvl}.

\item[(\uppercase\expandafter{\romannumeral2})] In Scheme II, including the $P_c(4380)$ discovered by LHCb in 2015~\footnote{Note that the broad $P_c(4380)$ with a large decay width has not been confirmed in later LHCb experiments as mentioned in the Introduction \cite{Roca:2016tdh}. Nevertheless, some studies have reproduced the $P_c(4380)$ mass through different methods, such as the production rates analysis of pentaquarks \cite{Du:2019pij} and mass spectrum prediction of the seven $P_c$ multiplets \cite{Liu:2019-prl}. }, we now have four masses of the $P_c(4312)$, $P_c(4440)$, $P_c(4457)$, and $P_c(4380)$ states, thus the four low-energy couplings could be determined, allowing us to derive the effective range $r_0$ for the $\bar{D}^{(*)}\Sigma_c^{(*)}$ molecules. In this way, we can check whether including the $P_c(4380)$ mass predicts the expected large $a_0\gg1/m_{ex}$ and natural $r_0\sim1/m_{ex}$, as suggested by Eqs.~\eqref{eq:Weinberg compositeness for a_0} and \eqref{eq:Weinberg compositeness} for the conjected molecular picture of the pentaquarks, thereby providing insight into the existence of the $P_c(4380)$ state.

\item[(\uppercase\expandafter{\romannumeral3})] In Scheme III, the suppression of the NLO low-energy couplings from the power counting can be utilized to estimate the $D_b$ term, allowing to fix four couplings with three inputs.  

According to Eqs.~\eqref{eq: NLO-1}-\eqref{eq: NLO-3} and considering that the NLO $D_b$ term might be suppressed around $\Lambda^{2}$ compared to the LO couplings (see Eqs.~\eqref{eq: R-value} and \eqref{eq:KSW power counting}), we can vary $D_b$ within a range of 
\begin{eqnarray}
   D_b \sim N \times \, \frac{1}{\Lambda^{2}} \times [-C, \, C] \, ,
   \label{eq:Db variation from EFT}
\end{eqnarray} 
where $N \sim 10$ is the magnitude chosen to ensure that the $D_b$ contribution can be fully considered, and $C \sim 3 \, \mathrm{fm}^{2}$ is the value of the LO couplings $C_0$ of $C_a$, $C_a-\frac{4}{3}C_b$, and $C_a+\frac{2}{3}C_b$ for $P_c(4312)$, $P_c(4440)$, and $P_c(4457)$. 

Given that $D_b$ has been determined within the above-specified range, the other three couplings of $C_a$, $C_b$, and $D_a$ could be fixed by solving the Lippmann-Schwinger equation for the $\bar{D}^{(*)}\Sigma_c$ systems to reproduce the $P_c(4312)$, $P_c(4440)$, and $P_c(4457)$ masses. 
   
\end{itemize}

\section{Numerical results}\label{sec: 6-results}

As in Ref.~\cite{Liu:2019-prl}, we are going to distinguish two scenarios: scenario A corresponds to $P_c(4440)$ having $J^P=\frac{1}{2}^-$ and $P_c(4457)$ having $J^P=\frac{3}{2}^-$, and scenario B represents the alternative assignment.   

By checking whether the results for scenario A or B can match the pure molecular picture of large, unnatural scattering length $a_0 \gg 1/m_{ex}$ and natural effective range $r_0\sim 1/m_{ex}$ as suggested by Eqs.~(\ref{eq:Weinberg compositeness for a_0}) and \eqref{eq:Weinberg compositeness}, one hopes to identify the spins of  $P_{c}(4440)$ and $P_{c}(4457)$. In addition to $a_0$ and $r_0$, another parameter $R$ can be defined as below for comparison 
\begin{eqnarray}
    R=\Lambda^2 \frac{C_2}{C_0} \, ,
    \label{eq:R parameter}
\end{eqnarray}
where the $C_2$ is the sum of $D_a$ and $D_b$, and $C_0$ is the the sum of $C_a$ and $C_b$. The natural value for this $|R|$, calculated from the contact effective field theory we constructed above, is expected to be around $R_{\rm natu.} = 1/\alpha \simeq 2.5$ from the power counting given in Eq.~(\ref{eq:KSW power counting}).
 
\subsection{Results for Scheme \uppercase\expandafter{\romannumeral1}}

First, we show the predictions of Scheme \uppercase\expandafter{\romannumeral1} where the $D_b$ coupling, representing the NLO spin-spin interaction term, is simply neglected in Eqs.~\eqref{eq: NLO-1}-\eqref{eq: NLO-3}. Taking the masses  of the $P_c(4312)$, $P_c(4440)$, and $P_c(4457)$ states as inputs, the resulting scattering length $a_0$ and effective range $r_0$ for these three pentaquarks are summarized in Table~\ref{tab:$r_0$ with scheme A}.

\begin{table*}[ht!]
\setlength{\tabcolsep}{8pt}
\caption{\justifying{The scattering length $a_0$,  effective range $r_0$, and parameter $|R|$ for $P_c(4312)$, $P_c(4440)$, and $P_c(4457)$ obtained in scheme~\uppercase\expandafter{\romannumeral1}, where the $D_b$ term has been neglected and two cutoffs of $0.5 \, \mathrm{GeV}$ and $1 \, \mathrm{GeV}$ are used for $\Lambda$. The $P_c(4312)$, $P_c(4440)$, and $P_c(4457)$ states are referred to $P_{c1}$, $P_{c2}$, and $P_{c3}$, respectively.}}
\label{tab:$r_0$ with scheme A}
\begin{tabular}{cc|ccc|ccc|ccc}
\hline\hline
   \multirow{2}{*}{Scenario}   &  \multirow{2}{*}{$\Lambda (\mathrm{GeV})$}   & \multicolumn{3}{c|}{$a_{0} (\mathrm{fm})$} & \multicolumn{3}{c|}{$r_{0} (\mathrm{fm})$}  & \multicolumn{3}{c}{$|R|$} \\ 
     \cline{3-11}
    & & $P_{c1}$ & $P_{c2}$ & $P_{c3}$ &$P_{c1}$ & $P_{c2}$ & $P_{c3}$ & $P_{c1}$ & $P_{c2}$ & $P_{c3}$ \\
\hline\hline
 \multirow{2}{*}{A} & 0.5 & 2.09 & 1.58 & 2.57 & $-0.04$ & $-0.02$ & $-0.06$ & 0.3 & 0.2 & 0.3 \\
  & 1 & 1.42 & $-0.34$ & $-0.07$ & $-0.54$ & $-0.31$ & $-0.72$ & 5.3 & 4.1 & 6.2 \\
\hline\hline
\multirow{2}{*}{B} & 0.5 & 2.88 & 2.45 & 3.40 & 0.61 & 0.53 & 0.75 & 3.0 & 3.4 & 2.8 \\  
  & 1 & 2.24 & 1.74 & 2.75 & 0.50 & 0.44 & 0.58 & 4.1 & 4.5 & 4.0 \\
\hline
\end{tabular}

\end{table*}

The effective range $r_0$ of $P_c(4440)$ and $P_c(4457)$ are positive in scenario B but negative in scenario A. This result indicates that, in the molecule picture, it is more natural to identify $P_c(4440)$ as the $J^P=\frac{3}{2}^-$ $\bar{D}^* \Sigma_c$ molecule and $P_c(4457)$ as the $J^P=\frac{1}{2}^-$ one. Additionally, the positive effective ranges derived for Scenario B fall within the range of $[0.4, 0.75] \, \mathrm{fm}$, which is indeed of the order of $\mathcal{O}(1/{m_{ex}})$, assuming the mass $m_{ex}$ of the exchanged meson mediating the $\bar{D}^{(*)}\Sigma_c$ interaction is approximately $500 \, \mathrm{MeV}$. This is close to the masses of mesons such as $\sigma$, $\rho$, $\omega$, which are often considered in the context of the one-boson exchange model for the $\bar{D}^{(*)}\Sigma_c$ molecules. These results for $r_0$ are consistent with the conclusion from the Weinberg compositeness criterion in Eq.~(\ref{eq:Weinberg compositeness}) that for pure molecular states with $Z=0$, $r_0$ should be positive and of the order of $\mathcal{O}(1/{m_{ex}})$.

Furthermore, the large scattering length $a_0\gg 1/ m_{ex}$ predicted for scenario B is also around $1/\gamma$ as expected from Eq.~(\ref{eq:Weinberg compositeness for a_0}), where $\gamma =\sqrt{2\mu E_B}\sim 100-200$ $\rm{MeV}$ are the binding momenta of $P_c(4312)$, $P_c(4440)$, and $P_c(4457)$ in the $\bar{D}^{(*)}\Sigma_c$ molecular picture.

\subsection{Results for Scheme \uppercase\expandafter{\romannumeral2}}

\begin{table*}[ht!]
\setlength{\tabcolsep}{8pt}
\caption{\justifying{The scattering length $a_0$, effective range $r_0$, and parameter $|R|$ for $P_c(4312)$, $P_c(4380)$, $P_c(4440)$, and $P_c(4457)$ obtained in scheme~\uppercase\expandafter{\romannumeral2} with two cutoffs of $0.5 \, \mathrm{GeV}$ and $1 \, \mathrm{GeV}$ are used for $\Lambda$. The $P_c(4312)$, $P_c(4440)$, $P_c(4457)$, and $P_c(4380)$ are referred to as $P_{c1}$, $P_{c2}$, $P_{c3}$, and $P_{c4}$, respectively.}}
\label{tab:$r_0$ with scheme B}
\begin{tabular}{cc|cccc|cccc|cccc}
\hline\hline
     \multirow{2}{*}{Scenario}   &  \multirow{2}{*}{$\Lambda (\mathrm{GeV})$}   & \multicolumn{4}{c|}{$a_{0} (\mathrm{fm})$} & \multicolumn{4}{c|}{$r_{0} (\mathrm{fm})$}  & \multicolumn{4}{c}{$|R|$} \\ 
     \cline{3-14}
    & & $P_{c1}$ & $P_{c2}$ & $P_{c3}$ & $P_{c4}$ & $P_{c1}$ & $P_{c2}$ & $P_{c3}$ & $P_{c4}$ & $P_{c1}$ & $P_{c2}$ & $P_{c3}$ & $P_{c4}$ \\
\hline\hline
\multirow{2}{*}{A} & 0.5 &5.03 & 2.88 & 6.37 & 4.81 & 1.68 & 0.75 & 2.28 & 1.66 & 6.7 & 3.6 & 8.6 & 6.7 \\
  & 1 & 5.01 & 2.07 & 7.00 & 4.55 & 1.87 & 0.65 & 2.62 & 1.85 & 13.2 & 5.5 & 17.9 & 13.2\\
\hline\hline
\multirow{2}{*}{B} & 0.5& 5.01 & 3.29 & 32.82 & 4.81 & 1.68 & 0.93 & 4.10  & 1.66 & 6.7 & 13.5 & 4.2 & 6.7\\ 
 & 1 & 5.01 & 2.54 & $-27.30$ & 4.55 
& 1.87 & 0.89 & 4.62 & 1.85 & 13.2 & 28.9 & 7.0 & 13.2\\ 
\hline
\end{tabular}
\end{table*}

In Scheme \uppercase\expandafter{\romannumeral2}, the $P_c(4380)$ is considered as an additional input alongside $P_c(4312)$, $P_c(4440)$, and $P_c(4457)$. With the masses of these four pentaquark states, the four couplings appearing in Eqs.~(\ref{eq: NLO-1})-\eqref{eq: NLO-3} can now all be fixed. The effective ranges for $P_c(4380)$, $P_c(4312)$, $P_c(4440)$, and $P_c(4457)$ are all positive for either scenario A or B, suggesting that both are consistent with the molecular picture. However, among the predicted positive $r_0$ for these four pentaquarks, there are always three with a magnitude larger than the natural value of order $\mathcal{O}(1/{m_{ex}})$, which does not fit well with the molecular picture.

The ratios $R$ in Eq.~(\ref{eq:R parameter}) for the four $P_c$ states, listed in Table \ref{tab:$r_0$ with scheme B}, are often 10-30 times greater than $\Lambda^{-2}$ for both scenarios A and B, which deviate significantly from the expectation of $1/\alpha\simeq2.5$ as required by the power counting in Eqs.~(\ref{eq:KSW power counting}) and (\ref{eq:R parameter}). 

These abnormal $a_0$, $r_0$, and $R$ obtained for both spin assignments suggest that $P_c(4380)$ might not be considered a molecular state together with $P_c(4312)$, $P_c(4440)$, and $P_c(4457)$. The $P_c(4380)$ may be excluded as an HQSS molecular partner of $P_c$ states. However, we will demonstrate below that the classification of $P_c(4380)$ as the HQSS partner is sensitive to its mass, where the broad $P_c(4380)$ discovered in 2015 cannot yet be ruled out because its mass, predicted from scenario B to be around $\sim 4376 \, \rm{MeV}$, is within the experimental uncertainty range for the $P_c(4380)$ mass.

\subsection{Results for Scheme \uppercase\expandafter{\romannumeral3}}
Introducing the variation in Eq.~(\ref{eq:Db variation from EFT}) based on the EFT power counting, the $D_b$ coupling can vary within a reasonable range. For each $D_b$ value within this range, the other three low-energy coupling of $C_a$, $C_b$, and $D_a$ can be fixed by identifying the $\mathcal{T}$ poles in Eq.~\eqref{eq. T=V+VGT} as $P_c(4312)$, $P_c(4440)$, and $P_c(4457)$ for $\bar{D}^{(*)}\Sigma_c$ molecules. 

\begin{figure*}[htbp]
    \centering
    \includegraphics[width=\textwidth]{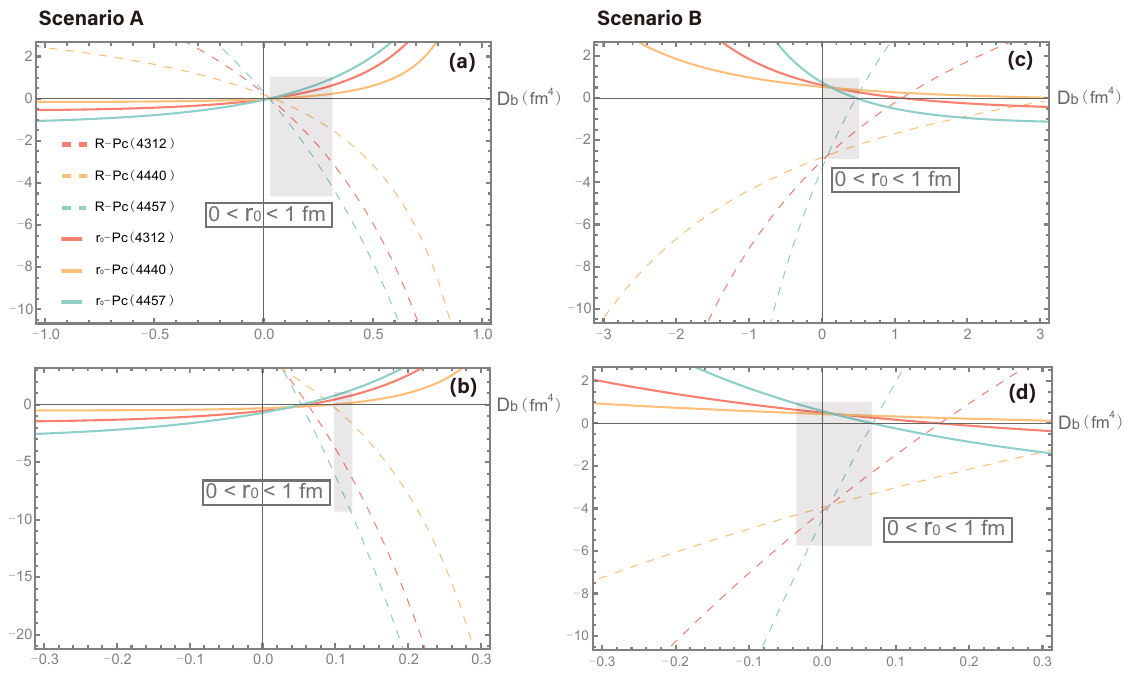}
    \caption{\justifying{The ratio $R$ and effective range $r_0$ of $P_c(4312)$, $P_c(4440)$, and $P_c(4457)$ obtained for Scheme \uppercase\expandafter{\romannumeral3}. Panels (a) and (b) show the $R$ and $r_0$ results for scenario A with two cutoffs of $\Lambda=0.5 \, \mathrm{GeV}$ and $\Lambda=1 \, \mathrm{GeV}$, while panels  (c) and (d) show the $R$ and $r_0$ results for scenario B with two cutoffs of $\Lambda=0.5 \,\mathrm{GeV}$ and $\Lambda=1 \,\mathrm{GeV}$.}}
    \label{fig:r0-R-Db-plot}
\end{figure*}

In Fig.~\ref{fig:r0-R-Db-plot}, we present the variations of $R$ and $r_0$ of $P_c(4312)$, $P_c(4440)$, and $P_c(4457)$ as functions of $D_b$. For scenario A [see Figs.~\ref{fig:r0-R-Db-plot}(a-b)], for both cutoffs of $\Lambda=0.5 \, \mathrm{GeV}$ and $\Lambda=1 \, \mathrm{GeV}$, the effective range $r_0$ can not simultaneously have a natural positive value around $\mathcal{O}(1/m_{ex}) \sim 0.5 \, \mathrm{fm}$ for $P_c(4312)$, $P_c(4440)$, and $P_c(4457)$. In addition, the ratio $R$ for these pentaquarks exceeds the natural range of $\mathcal{O}(1/\alpha)\simeq2.5$ for $r_0$ around $\mathcal{O}(1/m_{ex})$ (shaded area with $0<r_0<1 \, \mathrm{fm}$ in Fig.~\ref{fig:r0-R-Db-plot}). 

However, for scenario B [see Figs.~\ref{fig:r0-R-Db-plot}(c-d)], the same positive effective range $r_0$ around $\mathcal{O}(1/m_{ex}) \sim 0.5 \, \mathrm{fm}$ for $P_c(4312)$, $P_c(4440)$, and $P_c(4457)$ can be obtained for a natural value of $|R| \sim 1-3$. This result corroborates again the conclusion obtained for Scheme \uppercase\expandafter{\romannumeral1}, indicating that in the HQSS framework, $P_c(4440)$ should have spin-parity $J^P=\frac{3}{2}^-$ and $P_c(4457)$ spin-parity $J^P=\frac{1}{2}^-$. 

\begin{figure*}[htbp]
    \centering
    \includegraphics[width=\textwidth]{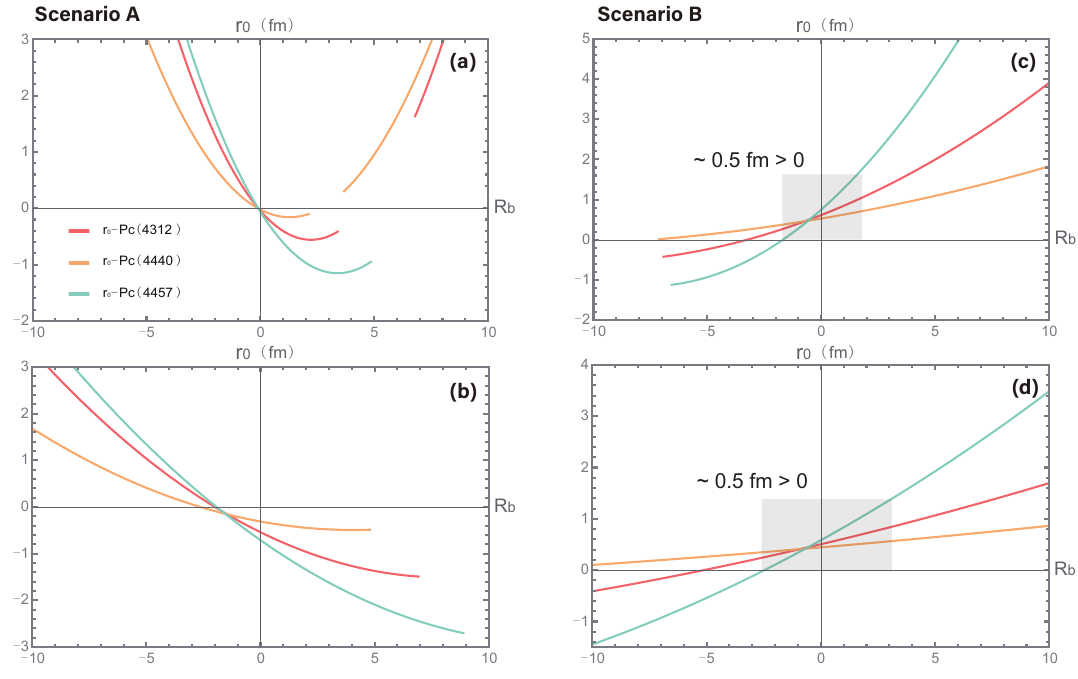}
    \caption{\justifying{The variation of the effective range $r_0$ with the ratio $R_b$ for $P_c(4312)$, $P_c(4440)$, and $P_c(4457)$ obtained for Scheme \uppercase\expandafter{\romannumeral3}. Panels (a) and (b) show the $r_0$ results for scenario A with two cutoffs of $\Lambda=0.5 \, \mathrm{GeV}$ and $\Lambda=1 \, \mathrm{GeV}$, while panels (c) and (d) denote the $r_0$ results for scenario B with two cutoffs of $\Lambda=0.5 \, \mathrm{GeV}$ and $\Lambda=1 \, \mathrm{GeV}$.}}
     \label{fig:r0-Rb-plot}
\end{figure*}

We also study the relation between the predicted effective range $r_0$ and the value of $R_b=\Lambda^2\frac{D_b}{C_0}$ to check further the naturalness of the contact EFT we employed. Only the NLO $D_b$ coupling is chosen to study the power counting. $C_0$ corresponds to $C_a$, $C_a-\frac{4}{3}C_b$, and $C_a+\frac{2}{3}C_b$ for the three $P_c$ states, respectively. Fig.~\ref{fig:r0-Rb-plot} shows that the effective ranges $r_0$ of $P_c(4312)$, $P_c(4440)$, and $P_c(4457)$ in scenario A are negative for $R_b$ around the natural range of $[-3,3]$. However, in scenario B, while $R_b$ varies within $[-3,3]$, the effective ranges $r_0$ for $P_c(4312)$, $P_c(4440)$, and $P_c(4457)$ are of natural positive values of $\mathcal{O}(1/{m_{ex}})\sim 0.5 \, \mathrm{fm}$, consistent with Eq.~(\ref{eq:Weinberg compositeness}). These results are also consistent with the conclusion drawn from Fig.~\ref{fig:r0-R-Db-plot}, indicating that it is more natural to consider $P_c(4440)$ as the $J^P=\frac{3}{2}^-$ and $P_c(4457)$ as the $J^P=\frac{1}{2}^-$ $\bar{D}^* \Sigma_c$ states if $P_c(4312)$, $P_c(4440)$, and $P_c(4457)$ are  $\bar{D}^{(*)} \Sigma_c$  heavy-quark symmetry partners in the molecular picture. 

From the condition where positive $r_0 \sim \mathcal{O}(1/m_{ex})$ of the three $P_c$ pentaquarks simultaneously emerge together with the natural $R\sim 2.5$ in scenario B, the $D_b\sim 0.15 \ \rm{fm^{4}}$ and $\sim 0.03 \ \rm{fm^{4}}$ can be determined for $\Lambda=0.5 \ \rm{GeV}$ and $\Lambda=1 \ \rm{GeV}$, respectively [see Figs.~\ref{fig:r0-R-Db-plot} (c-d)]. The above $D_b$ couplings can predict whether the broad $P_c(4380)$ state could be regarded as a $\bar{D} \Sigma_c^*$ molecule. From the results shown in Table \ref{tab:prediction of Pc-4380 with scenario B}, the  $P_c(4380)$ mass is predicted as follows
\begin{eqnarray}
    m_{P_c(4380)} &=& 4375.84 \ \ \rm{MeV} \, \, (\Lambda=0.5 \ \rm{GeV}) \, ,\\
     m_{P_c(4380)} &=& 4375.6 \ \ \rm{MeV} \, \, \, (\Lambda=1 \ \rm{GeV}) \, .
\end{eqnarray}
These values are approximately $ 4 \, \rm{MeV}$ below the central mass value of the $P_c(4380)$ state as observed in 2015, but they remain comfortably within the experimental uncertainties \cite{LHCb:2015yax}. Therefore, the broad $P_c(4380)$ cannot yet be excluded as the HQSS molecular partner of $P_c(4312)$, $P_c(4440)$, and $P_c(4457)$; meanwhile the predicted mass of $P_c(4380)$ is constrained to $\sim 4376 \, \rm{MeV}$ based on the naturalness of the power counting of the contact effective field theory.
 
\begin{table*}[ht!]
\setlength{\tabcolsep}{8pt}
\caption{\justifying{With $D_b\sim 0.15 \ \rm{fm^{4}}$ and $\sim 0.03 \ \rm{fm^{4}}$  for $\Lambda=0.5 \ \rm{GeV}$ and $\Lambda=1 \ \rm{GeV}$ from scenario B, the mass of $\bar{D} \Sigma_c^*$, the scattering length $a_0$, the effective range $r_0$, and the parameter $R$ for $P_c(4312)$, $P_c(4380)$, $P_c(4440)$, and $P_c(4457)$ are predicted.}}
\label{tab:prediction of Pc-4380 with scenario B}
     \begin{tabular}{cc|cccc|cccc|cccc}
\hline\hline
       \multirow{2}{*}{$\Lambda (\mathrm{GeV})$} & \multirow{2}{*}{$m_{\bar{D} \Sigma_c^*} (\mathrm{MeV})$}  & \multicolumn{4}{c|}{$a_{0} (\mathrm{fm})$} & \multicolumn{4}{c|}{$r_{0} (\mathrm{fm})$}  & \multicolumn{4}{c}{$|R|$} \\ 
     \cline{3-14}
    & & $P_{c1}$ & $P_{c2}$ & $P_{c3}$ & $P_{c4}$ & $P_{c1}$ & $P_{c2}$ & $P_{c3}$ & $P_{c4}$ & $P_{c1}$ & $P_{c2}$ & $P_{c3}$ & $P_{c4}$ \\
\hline\hline
 0.5 & 4375.84 & 2.73 & 2.37 & 3.08 & 2.67 & 0.50 & 0.48 & 0.46 & 0.49 & 2.50 & 2.63 & 2.19 & 2.50 \\
 1 & 4375.6 & 2.13 & 1.69 & 2.52 & 2.06 & 0.40 & 0.41 & 0.31  & 0.39 & 3.31 & 3.67 & 2.49 & 3.31\\  
\hline
\end{tabular}

\end{table*}

\section{Discussion and Conclusion}\label{sec: 7-conclusions}

 This study investigated the effective ranges of the recently discovered $P_c(4312)$, $P_c(4440)$, and $P_c(4457)$ states in the molecular picture and the contact effective field theory framework. The Weinberg compositeness criterion indicates for a pure hadronic molecular state, the effective range $r_0$ should be positive and of the order of $\mathcal{O}(1/{m_{ex}})$, while the scattering length is unnaturally large, i.e., $a_0 \gg \mathcal{O}(1/{m_{ex}})$.

Taking into account two alternative spin assignments for $P_c(4440)$ and $P_c(4457)$, we proposed three schemes for the NLO contact potentials. This allowed us to determine the scattering length $a_0$, effective range $r_0$, and the parameter $R$, where $R$ is quantified through a naturalness criterion dictated by the power counting, for the $P_c(4312)$, $P_c(4440)$, and $P_c(4457)$ pentaquarks. 

We showed that only scenario B can accommodate simultaneously a positive effective range $r_0 \sim \mathcal{O}(1/m_{ex})$, a large scattering length $a_0\gg\mathcal{O}(1/m_{ex})$, and a natural $R\sim2.5$. As a result, it seems to be more natural to assign $J^P=\frac{3}{2}^-$ to $P_c(4440)$ and $J^P=\frac{1}{2}^-$ to $P_c(4457)$ in the molecular $\bar{D}^* \Sigma_c$ picture. This conclusion agrees with Refs.~\cite{Liu:2019zvb,Yalikun:2021bfm,Du:2019pij,Du:2021fmf,Yamaguchi:2019seo} and indicates that within the molecular states composed of the same components, the pentaquarks with higher spin prefer a lower mass. This is consistent with scenario B of Ref. \cite{Liu:2019-prl}. 

It should be noted that our study is performed in single channel. Coupled-channel effects, such as $\bar{D} \Lambda_{c1}(2595)$~\cite{Geng:2017hxc,Burns:2022uiv}, might alter the aforementioned conclusion. This is because the threshold for $\bar{D} \Lambda_{c1}(2595)$ is close to $P_c(4440)$ and $P_c(4457)$, and it may couple with $\bar{D}^* \Sigma_c$. However, limited by the available mass inputs, determining the potential of $\bar{D} \Lambda_{c1}(2595)-\bar{D} \Lambda_{c1}(2595)$ and crossed-channel $\bar{D} \Lambda_{c1}(2595)-\bar{D}^* \Sigma_c$ is challenging. Consequently, studying the effective range within coupled-channel are beyond the scope of our current work. Additionally, we assumed $Z=0$ to obtain $r_0 > 0$. Therefore, if $P_{c}(4312)$, $P_{c}(4440)$ and $P_{c}(4457)$ contain a significant compact component, the above conclusion of positive $r_0$ might change. Nevertheless, the range of $r_0$ being estimated as $|r_0|\sim\mathcal{O}(1/m_{ex})$ should still be satisfied for dominantly molecular states.

Moreover, since the LHCb data of 2019 can be fitted equally well by models that either include or exclude the Breit-Wigner contribution from $P_c(4380)$, the potential role of $P_c(4380)$ as a HQSS molecule, akin to $P_c(4312)$, $P_c(4440)$, and $P_c(4457)$, remains uncertain. Our results for Scheme \uppercase\expandafter{\romannumeral2} and \uppercase\expandafter{\romannumeral3} suggest that if the $P_c(4380)$ is to be regarded as the HQSS molecular partner of $\bar{D} \Sigma_c^*$ alongside $P_c(4312)$, $P_c(4440)$, and $P_c(4457)$, its mass should be constrained to around $ 4376 \, \rm{MeV}$ based on the naturalness argument from the power counting, which resides roughly $4 \, \rm{MeV}$ below the experimental central mass of $P_c(4380)$. 

These findings about discriminating the spins of $P_c$ states and constraining the hidden-charm pentaquark predictions are relevant for future experimental investigations and further theoretical studies on the inner structure of pentaquarks.

\begin{acknowledgments}

We would like to thank Mao-Jun Yan, Jun-Xu Lu, Eulogio Oset and Manuel Pavon Valderrama for the helpful discussions. This work is partly supported by the National Key R\&D Program of China under Grant No. 2023YFA1606703, the Postdoctoral Fellowship Program of CPSF under Grant Number GZC20241765, the Postdoctoral Fellowship Program of Gansu Province, and by the National Natural Science Foundation of China under Grant Nos. 12075288, 12435007 and 12361141819. It is also supported by the Youth Innovation Promotion Association CAS.

\end{acknowledgments}

\bibliography{reference}

\end{document}